\documentclass[twoside,reqno]{HERON}
\usepackage{epsfig,cite,colordvi}
\usepackage{rotating}
\usepackage{url}
\usepackage{graphicx}% Include figure files
\usepackage{dcolumn}% Align table columns on decimal point
\usepackage{bm}% bold math
\usepackage{epsfig}
\usepackage{amsmath}
\usepackage{makeidx}
\def\be{\begin{equation}}

\def\ee{\end{equation}}
\def\barr{\begin{array}}
\def\earr{\end{array}}

\def\nn8{\nonumber\\[2pt]}
\def\l{\left}
\def\r{\right}
\def\dis{\displaystyle}
\def\ed{\end{document}}

\def\chsu4{SO_{sdST}(36) \supset SO_{sST}(6) \oplus SO_{dST}(30)}

\begin{document}

\title{Deformed shell model study of heavy N=Z nuclei and dark matter detection}

\runningheads{Deformed shell model studies}{R. Sahu, V.K.B. Kota}

\begin{start}

\author{R. Sahu}{1}, \coauthor{V.K.B. Kota}{2}

\index{Sahu, R.}
\index{Kota, V.K.B.}

\address{
Physics Department,
Berhampur University,
Berhampur-760007,
Odisha, India.}{1}

\address{
Physical Research Laboratory,
Ahmedabad-380009,
India.}{2}

\begin{Abstract}

Deformed shell model (DSM) based on Hartree-Fock intrinsic states is applied to
address two current problems of interest. Firstly, in  the $f_{5/2}pg_{9/2}$
model space  with jj44b effective interaction along with isospin projection, DSM
is used to describe the structure of the recently observed low-lying  $T=0$ and
$T=1$ bands in the heavy odd-odd N=Z nucleus $^{66}$As. DSM results are close to
the data and also to the shell model results. For the $T=1$ band, DSM predicts
structural change at  $8^+$ just as in the shell model. In addition, the lowest
two $T=0$  bands are  found to have quasi-deuteron structure above a $^{64}$Ge
core and the $5^+$ and $9^+$ levels of the third $T=0$ band are found to be
isomeric states. Secondly, in a first application of DSM to dark matter,
detection rates for the lightest supersymmetric particle (a dark matter
candidate) are calculated with $^{73}$Ge as the detector.

\end{Abstract}

\end{start}

\section{Introduction}

There has been considerable interest in investigating the structure of the
nuclei in the mass region $A = 60-100$ and in particular odd-odd $N = Z$ nuclei
as these nuclei are expected to give new insights into neutron-proton ($np$)
correlations that  are hitherto unknown.   The N=Z nuclei in this mass region
lie near the proton drip-line. With the development of radioactive ion beam 
facilities and large detector arrays, new experimental results for the energy
spectra of $^{62}$Ga \cite{david}, $^{66}$As \cite{66as}, $^{70}$Br \cite{70br},
$^{74}$Rb \cite{Ru-96}, $^{78}$Y \cite{exp-4}, $^{82}$Nb \cite{Garns-08} and 
$^{86}$Tc \cite{Garns-08} have opened up challenges in developing models for
describing and predicting the spectroscopic properties of these nuclei. On the
other hand, many interesting phenomena  have been observed with shape changes
and delayed alignments in even-even $N = Z$ nuclei from $^{64}$Ge to $^{88}$Ru.
For example, $^{64}$Ge exhibits $\gamma$-soft structure \cite{Sta-07}, $^{68}$Se
exhibits oblate shape in the ground state \cite{68se}, $^{72}$Kr
\cite{72kr1,72kr2,72kr3} exhibits shape coexistence, $^{76}$Sr and $^{80}$Zr
have large ground state deformations \cite{76Sr,Lister-87} and so on.  Recently,
evidence for a spin-aligned $np$ isoscalar paired phase has been reported from
the level structure of $^{92}$Pd \cite{cederwall}.  Also many even-even $N = Z$
nuclei in this region are waiting point nuclei for rp-process  nucleosynthesis
\cite{Fuller} and hence they are of astrophysical interest.  The recent
development of the recoil-$\beta$-tagging technique provides a tool to study
medium-mass nuclei around the $N=Z$ line. In Jyv\"{a}skyl\"{a}, excited states
of $^{66}$As were populated using $^{40}$Ca($^{28}$Si,$pn$)$^{66}$As 
fusion-evaporation reaction at beam energies of 83 MeV and 75 MeV. Also in this
experiment half-lives and ordering of the two known isomeric states ($5^+$ at 
1354 keV and $9^+$ at 3021 keV) have been determined with improved accuracy
\cite{66as}.

The deformed shell model (DSM), based on Hartree-Fock (HF) deformed intrinsic
states with angular momentum projection and band mixing, is established to be a
good model to describe the properties of nuclei in the mass range A=60-100.
Also, for N=Z odd-odd nuclei, methods for isospin projection within DSM are 
developed and applied. See \cite{ks-book} for details regarding DSM and it is
found to be quite  successful in describing  spectroscopic properties, double
beta decay half-lives, $\mu -e$ conversion in the field of the nucleus and so
on. Following these, in the present paper presented are results of two
investigations using  DSM. In the first application, recent data \cite{66as} for
the  heavy N=Z nucleus $^{66}$As are analyzed using DSM with isospin projection.
Secondly,  DSM is employed to calculate the detection rates for the lightest
supersymmetric particle (a dark matter candidate) with $^{73}$Ge as the
detector. These are described in Sections 2 and 3 respectively. Finally Section
4 gives conclusions.  

\begin{table*}
\caption{Prolate (P) and oblate (O) intrinsic states (configurations) used for
$^{66}$As in the DSM calculation. For each of these, the total $K$ value and
isospin $T$ are given in the Table. Superscript $(2p,2n)$ implies that the orbit
is occupied by two protons and two neutrons and similarly the superscript
$(p,n)$. In addition, the superscripts $p(n)$ and $n(p)$ imply that the orbit(s)
is(are) alternatively occupied by  a proton and a neutron or a neutron and a
proton. Note that the entries with a $*$ correspond to four particle
configurations and those with $**$ correspond to six particle configurations.
Similarly, $p^-$ is proton hole and $n^-$ is  neutron hole. Note that
$X=(1/2^-)_1^{2p,2n}$ and $Y=(3/2^-)_1^{2p,2n}$ in the table. }
\begin{tabular}{c|c|c|c|l}
\hline 
\hline
No. & shape & $\;\;K\;\;$ & $\;\;T$  & $\;\;\;$Configuration \\  
\hline
1. & P & $3^+$ & $0$ &  $X\,(1/2^-)_2^{2p,2n}
(3/2^-\uparrow)_1^{p,n}$ \\
2,3. & P & $0^+$ & $0,1$ & $X\,(1/2^-)_2^{2p,2n} 
(3/2^- \uparrow)_1^{p(n)} (3/2^- \downarrow)_1^{p(n)}$ \\ 
4. & P & $1^+$ & $0$ &  $X\,(1/2^-)_2^{2p,2n} 
(1/2^+ \uparrow)_1^{p,n}$ \\
5,6. & P & $0^+$ & $0,1$ & $X\,(1/2^-)_2^{2p,2n} 
(1/2^+ \uparrow)_1^{p(n)} (1/2^+ \downarrow)_1^{p(n)}$ \\ 
7,8. & P & $2^+$ & $0,1$ &  $X\,(1/2^-)_2^{2p,2n} 
(3/2^+ \uparrow)_1^{p(n)} (1/2^+ \uparrow)_1^{n(p)}$ \\ 
9,10. & P & $1^+$ & $0,1$ &  $X\,(1/2^-)_2^{2p,2n} 
(3/2^+ \uparrow)_1^{p(n)} (1/2^+ \downarrow)_1^{n(p)}$ \\
11,12. & P & $3^+$ & $0,1$ & $X\,(1/2^-)_2^{2p,2n} 
(3/2^- \uparrow)_1^{p(n)} (3/2^- \uparrow)_2^{n(p)}$ \\
13,14. & P & $0^+$ & $0,1$ & $X\,(1/2^-)_2^{2p,2n} 
(3/2^- \uparrow)_1^{p(n)} (3/2^- \downarrow)_2^{n(p)}$ \\
15,16. & P & $0^+$ & $0,1$ & $X\,(1/2^-)_2^{2p,2n} 
(3/2^- \uparrow)_2^{p(n)} (3/2^- \downarrow)_1^{n(p)}$ \\
17. & P & $1^+$ & $0$ & $X\,(3/2^-)_2^{2p,2n} 
(1/2^- \uparrow)_2^{p,n}$ \\
18,19. & P & $0^+$ & $0,1$ & $X\,(3/2^-)_2^{2p,2n} 
(1/2^- \uparrow)_2^{p(n)} (1/2^- \downarrow)_2^{p(n)}$ \\  
20. & P & $1^+$ & $0$ & $X\,(3/2^-)_2^{2p,2n} 
(1/2^+ \uparrow)_1^{p,n}$ \\
21,22. & P & $0^+$ & $0,1$ & $X\,(3/2^-)_2^{2p,2n} 
(1/2^+ \uparrow)_1^{p(n)} (1/2^+ \downarrow)_1^{p(n)}$ \\
23,24. & P & $1^+$ & $0,1$ & $X\,(1/2^- \downarrow)_2^{
p^-(n^-)} (3/2^-\uparrow)_1^{p^-(n^-)} $ \\
25,26. & P & $2^+$ & $0,1$ & $X\,(1/2^- \uparrow)_2^{
p^-(n^-)} (3/2^-\uparrow)_1^{p^-(n^-)} $ \\
27-32. & P$^*$ & $1^+$ & $0,1,2$ & $X\,(1/2^- \downarrow)_2^{
p^-(n^-)} [(3/2^-)_1 (3/2^- \uparrow)_2]^{p,n,p(n)}$ \\
33-38. & P$^*$ & $2^+$ & $0,1,2$ & $X\,(1/2^- \uparrow)_2^{p^-(n^-)} 
\l[(3/2^-)_1 (3/2^- \uparrow)_2\r]^{p,n,p(n)}$ \\ 
39-58. & P$^{**}$ & $0^+$ & $0,1,2,3$ & $X\,[(1/2^-)_2 (3/2^-)_1
(3/2^-)_2]^{3p,3n}$ \\
59-78. & P$^{**}$ & $0^+$ & $0,1,2,3$ & $X\, [(1/2^-)_2 (3/2^-)_1
(1/2^+)_1]^{3p,3n}$ \\
79. & O & $1^+$ & $0$ & $Y\,(5/2^-)_1^{2p,2n} 
(1/2^- \uparrow)_1^{p,n}$ \\
80,81. & O & $0^+$ & $0,1$ & $Y\,(5/2^-)_1^{2p,2n} 
(1/2^- \uparrow)_1^{p(n)} (1/2^- \downarrow)_1^{n(p)}$ \\
82. & O & $5^+$ & $0$ & $Y\,(1/2^-)_1^{2p,2n} 
(5/2^- \uparrow)_1^{p,n}$ \\
83,84 & O & $0^+$ & $0,1$ & $Y\,(1/2^-)_1^{2p,2n} 
(5/2^- \uparrow)_1^{p(n)} (5/2^- \downarrow)^{n(p)}$ \\
85. & O & $9^+$ & $0$ & $Y\,(1/2^-)_1^{2p,2n} 
(9/2^+ \uparrow)_1^{p,n}$ \\
86,87. & O & $0^+$ & $0,1$ & $Y\,(1/2^-)_1^{2p,2n} 
(9/2^+ \uparrow)_1^{p(n)} (9/2^+ \downarrow)^{n(p)}$ \\
88. & O & $9^+$ & $0$ & $Y\,(5/2^-)_1^{2p,2n} 
(9/2^+ \uparrow)_1^{p,n}$ \\
89,90. & O & $0^+$ & $0,1$ & $Y\,(1/2^-)_1^{2p,2n} 
(9/2^+ \uparrow)_1^{p(n)} (9/2^+ \downarrow)^{n(p)}$ \\
91-92. & O & $2^+$ & $0,1$ & $Y\,(1/2^- \downarrow)_2^{
p^-(n^-)} (5/2^-\uparrow)_1^{p^-(n^-)} $ \\
93,94. & O & $3^+$ & $0,1$ & $Y\,(1/2^- \uparrow)_2^{
p^-(n^-)} (5/2^-\uparrow)_1^{p^-(n^-)} $ \\
95-114. & O$^{**}$ & $0^+$ & $0,1,2,3$ & $Y\,[(1/2^-)_1 (5/2^-)_1 (9/2^+)_1]^{
3p,3n}$ \\
\hline\hline
\end{tabular}
\label{HF}
\end{table*}

\section{DSM results for spectroscopic properties of $^{66}$As}

\begin{figure*}
\includegraphics[width=5cm]{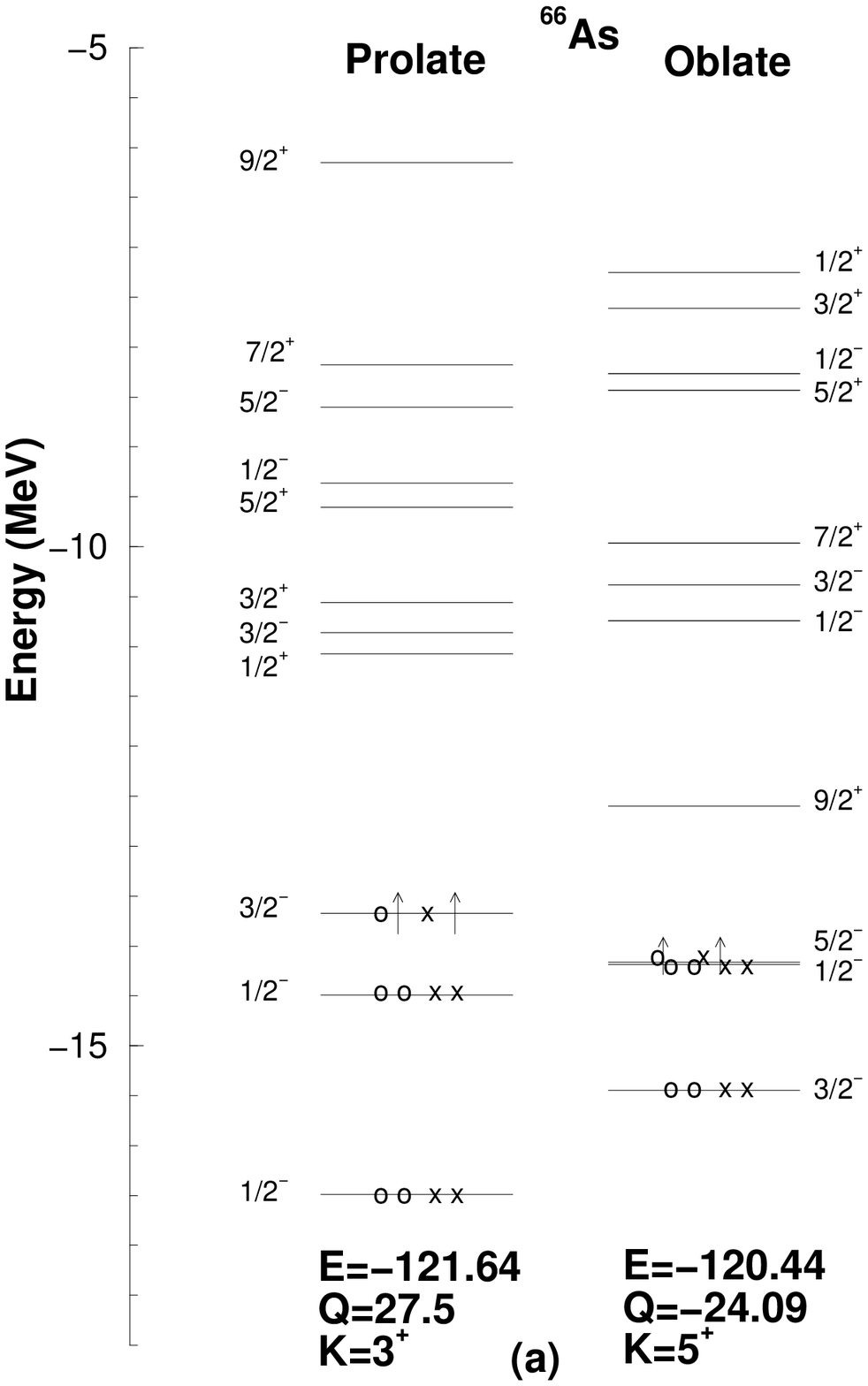} \vspace{8cm}
\includegraphics[width=5cm]{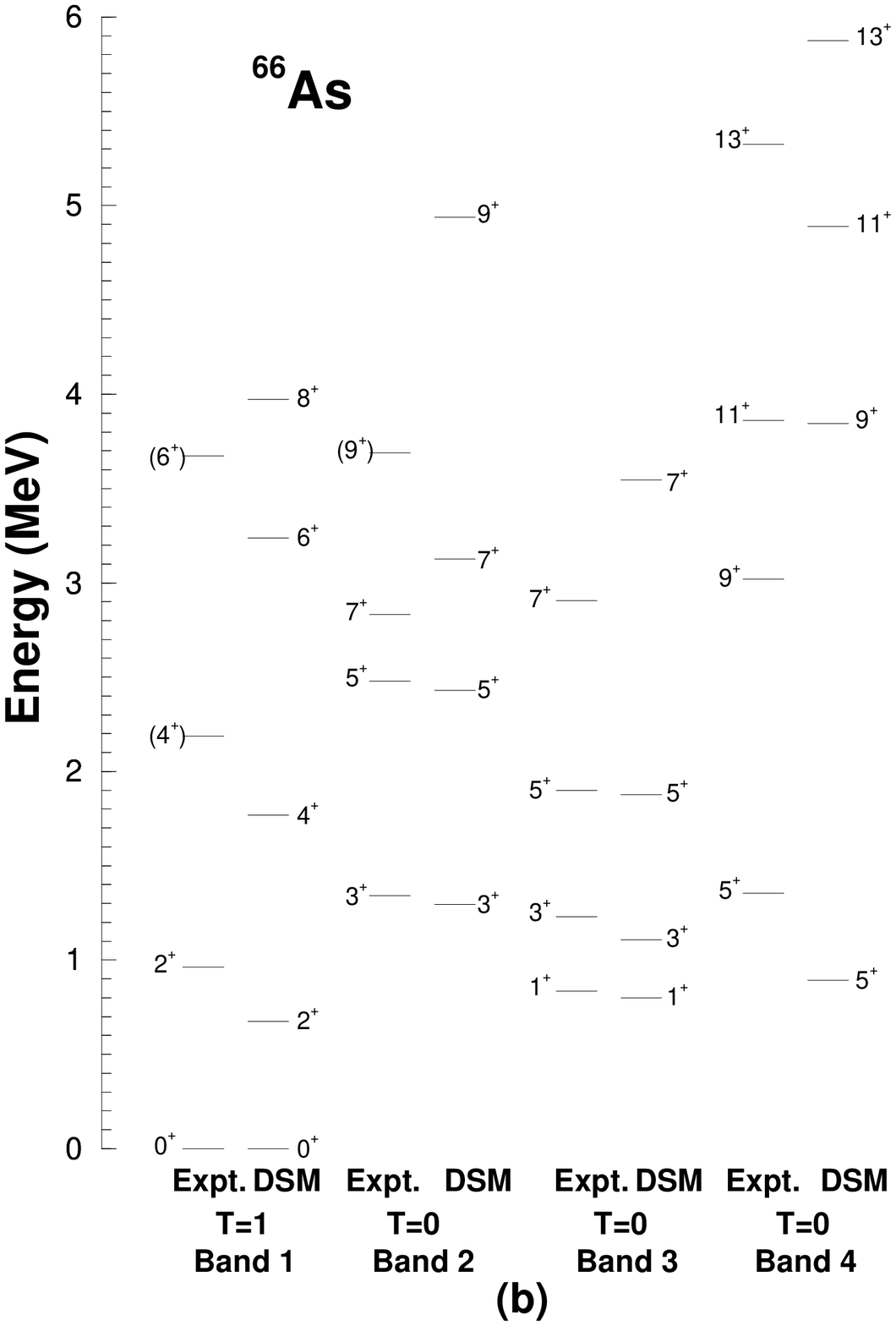}

\vskip -8cm

\caption{(a) The spectra of sp states for the lowest energy prolate and oblate
intrinsic HF configurations for  $^{66}$As. Protons are represented by circles
and the neutrons by crosses. The HF energy in MeV and the quadrupole moment in
the units of square of oscillator length parameter are also shown.  (b)
Comparison of deformed shell model results with experimental data \cite{66as}
for different bands with jj44b interaction. The band numbers shown in the figure
are according to Ref. \cite{66as}. Shell model results are obtained using
the same jj44b interaction by P.C. Srivastava for this nucleus \cite{pcs-1}. For
the levels in Band 1, with $J^\pi$ values in the same order as in the figure,
the SM energies are 0.0, 0.899, 2.273, 4.034 and 5.124 MeV respectively.
Similarly, for Band 2  members they are 0.889, 2.263, 3.585 and 4.163 MeV, for
Band 3 they are 0.426, 0.739, 1.568  and 2.972 MeV and for Band 4 they are
0.985, 1.958, 3.045 and 4.799 MeV respectively.}

\label{fig1}
\end{figure*}

Calculations are performed in the model space consisting of $2p_{3/2}$, 
$1f_{5/2}$, $2p_{1/2}$ and $1g_{9/2}$ orbits,  with $^{56}$Ni as the inert core.
For $^{66}$As nucleus, Fig.~\ref{fig1}(a) gives the HF single particle (sp)
spectrum (the states are labeled by $|k_\alpha \rangle$ where the $\alpha$ label
distinguishes different states with the same $k$ value)  for both prolate and
oblate solution obtained using the jj44b interaction given in \cite{jj44b}. 
The  prolate state is more bound compared to the oblate intrinsic state by more
than 1 MeV. In the prolate case, two protons and two neutrons occupy the lowest
$k=(1/2^-)_1$ sp state forming an alpha particle like structure and it has 
$T=0$. Similarly, the next $k=1/2^-$ state is filled by two protons and two
neutrons and the last unpaired proton and neutron occupying the lowest $k=3/2^-$
state. Then, the isospin of the nucleus is determined by these last proton and
neutron.   The total isospin for the configuration shown in Fig.~\ref{fig1}(a)
is $T=0$ as the odd proton and odd neutron, for $K=3^+$, form a  symmetric pair
in the $k$-space (here and elsewhere in this paper symmetry in $k$-space means
symmetry in space-spin co-ordinates as $k$ contains both space (orbital) and
spin co-ordinates). Particle-hole excitations over the lowest HF intrinsic state
(from both prolate and oblate solutions) generate excited HF intrinsic states.

By making particle-hole excitations for the six nucleons out side the lowest $k$
orbit, we have considered 114 configurations shown in Table~\ref{HF}.  
Out of these, 78 configurations are of prolate shape and 36 oblate
shape.  Since the lowest oblate configuration lies more than one MeV higher
compared to  the lowest prolate configuration, we have considered fewer oblate
configurations.  For the intrinsic states labeled 1-26 with prolate shape in 
Table ~\ref{HF}, the isospin is determined by the last proton and last neutron.
When they are in the same orbit with spin up, uniquely $T=0$. In other
situations, it is possible to have both symmetric and antisymmetric spatial
configurations giving $T=0$ and $T=1$ states respectively as shown in Table
~\ref{HF}. With oblate shape we have similar $np$ configurations for
intrinsic states labeled 79-94. In addition, with prolate shape a proton or a
neutron from the $(1/2^-)_2$ orbit can be promoted to the $(3/2^-)_2$ orbit.
Then we have a proton or neutron hole in the $(1/2^-)_2$ orbit. This excited
configuration is effectively a four particle (two proton and two neutron)
configuration giving six intrinsic states. Constructions of good isospin states
from these six states is described in \cite{SK-2005}. Projection of isospin from
the four particle configurations gives two $T=0$ states,  three $T=1$ states and
one $T=2$ state. States labeled 27-32 and 33-38 belong to this class. For
prolate case, we have also considered configurations with a $(3p,3n)$ system
distributed in six orbits. These six particle configurations  are labeled 39-58
and 59-78 in Table ~\ref{HF}. In this situation, distributing in all
possible ways the three protons and three neutrons will give twenty different
configurations and then isospin projection gives five $T=0$, nine $T=1$, five
$T=2$ and one $T=3$ intrinsic states.  The details regarding isospin projection
for the six particle case is given in Ref. \cite{KS-2009}. For oblate shape, we
have considered one set of six particle configurations and they are labeled
95-114 in Table ~\ref{HF}. Combining all these, we have a total of forty
four $T=0$ configurations and fifty $T=1$  configurations (in addition there are
seventeen $T=2$ configurations and  three $T=3$ configurations but they are not
relevant  as experimental data contains only $T=1$ and $T=0$ levels). We project
out good angular momentum states from different intrinsic states of a given
isospin and then perform band mixing calculations. The resulting spectrum for
$T=0$ and $T=1$ bands are compared with experiment in Fig.~\ref{fig1}(b). 

The DSM calculated  $T=1$ band agrees reasonably well with experiment.  Except
for the $2^+ \rightarrow 0^+$ separation,  the relative spacing of all other
levels are reasonably reproduced.  The $T=1$ levels up to $J=6^+$ mainly
originate from the lowest $T=1$ intrinsic state generated by the antisymmetric
combination of the configurations No. $(2,3)$ in Table 1.  Hence, there is no
change in the collectivity up to $J=6^+$. The shell model (SM) as well as the
DSM predicts the B(E2) values for the transition $8^+ \rightarrow 6^+$ to be
very small. For example, the B(E2) ratios  $B(E2,I \rightarrow I-2)$
$/B(E2,I-2\rightarrow I-4)$ with I=4, 6, 8 are 1.22, 0.97 and 0.001 in DSM. The
corresponding ratios for shell model are 1.29, 1.09 and 0.001.   The occupancy
of the $1g_{9/2}$ orbit obtained from SM does not change much up to spin $T=1,
J=6^+$ and is about 0.64 for both protons and neutrons. However, as we go to
$T=1, 8_1^+$ level, there is a dramatic change in the occupancy which is 1.05 in
SM. Thus, shell model predicts the structure of the $T=1, 8_1^+$ level to be
quite different from that of the other $T=1$ levels lying below. As a result,
the $B(E2)$ transition probability from $T=1, 8^+$ to $T=1, 6^+$ is small. This
is in agreement with the conclusion drawn from the DSM calculation which
predicts that the structure of the $T=1, 8_1^+$ level to be a quite different
from that of the $T=1,6^+_1$ level. This level originates from the $T=1$
projected intrinsic state in which three protons and three neutrons are
distributed in six single particle orbitals with configuration that corresponds
to No. 59-78 in Table ~\ref{HF}. This configuration has a proton and a neutron
in $g_{9/2}$ orbit just the occupancy given by SM. With the structure of the
$T=1, 8^+_1$ level being quite different from the $T=1,  6^+_1$, its E2
transition probability to $T=1, 6^+_1$ level is also small in DSM just as in SM.

Coming to the $T=0$ bands, it is seen from Fig. \ref{fig1}(b) that the DSM
calculated spectra for the first two $T=0$ bands (band 2 and band 3 in the
figure) agree reasonably well with experiment (also with shell model).    The
levels $3^+$, $5^+$, $7^+$ and $9^+$ of band 2 with $T=0$  mainly originate from
the lowest T=0 intrinsic state No. 1 given in Table 1. However, the mixing from
other intrinsic states increases  for the higher spin states.  The band 3 with 
$T=0$ consists of $1^+$, $3^+$, $5^+$ and $7^+$ levels. All these levels except
$7^+$ level are found to have similar structure. They mainly originate from the
lowest T=0 intrinsic state No. 1 in Table 1 and from the   symmetric combination
of the intrinsic states No. $(2,3)$ given in Table 1. Thus, both these two $T=0$
bands exhibit quasi-deuteron structure above a  $^{64}$Ge core.  

The band 4 (also with $T=0$) consists of a level with spin $5^+$ and in DSM this
level is essentially generated by the oblate configuration $(3/2^-)_1^{2p,2n}
(1/2^-)_1^{2p,2n} (5/2^-\uparrow)_1^{p,n}$ (No. 82 in  Table 1). The calculated
$B(E2)$'s from this level to the lower $3^+$ levels of band 2 and band 3 are
very small. Thus, this level is an isomeric state obtained from the totally
aligned $^1f_{5/2}$ $np$ configuration consistent with the claim
in \cite{66as}.  Also, the structure of this $5^+$ level is similar to the $7^+$
level of band 3. This is possibly the reason why in the experiment reported in
\cite{66as}, a large transition strength between these two levels is seen.  The
other levels of this band are $9^+$, $11^+$ and $13^+$. The $9^+$ energy from DSM
is higher than the experimental value by about 1 MeV.  However the relative
spacings are quite well reproduced. The $9^+$ level originates from the oblate
intrinsic state No. 85 given in Table 1 with configuration $(3/2^-)_1^{2p,2n}
(1/2^-)_1^{2p,2n} (9/2^+\uparrow)_1^{p,n}$. It has also strong mixing from the
prolate intrinsic states No. (5,6), (7,8) and (9,10) given in Table 1. The
calculated B(E2) values  from this level to the lower $7^+$ levels of bands 2
and 3 are very small. Thus, this level is  predicted to be a isomeric state with
totally aligned $np$ pair in $^1g_{9/2}$ orbit as the dominant
structure.   The yrast $11^+$ and $13^+$ levels in DSM do not have the same
structure and the levels with aligned structure appear at much higher energies
and they are not shown in the figure. Hasegawa et al \cite{Has-2005} have
performed a spherical shell model study using an extended pairing plus
quadrupole-quadrupole Hamiltonian and they also  found the $9^+$ levels to be a
isomeric state as in DSM. Cranked Nilsson-Strutinsky calculations and SM also
predict a band 4 with $9^+$, $11^+$ and $13^+$ (also higher $J^\pi$) with
totally aligned $pn$ pair in $^1g_{9/2}$ orbit \cite{pcs-1}. Thus, $^{66}$As
also shows spin aligned $np$ isoscalar pair phase as seen before in $^{92}$Pd
\cite{cederwall}.

It is important to add that the inclusion of six particle configurations on one
hand and mixing of several intrinsic states with  prolate and oblate shapes on
the other, are responsible for the good agreements seen for various bands
generated by DSM (the band 5 with three levels identified in experiment was  not
discussed here as the assignment of both the angular momentum and parity  of
these levels is uncertian). Also, the present calculations show considerable
improvements over the previous DSM results \cite{sk1} where only limited number
of two particle configurations are included and a realistic G-matrix interaction
with a phenomenologically adjusted monopole part as given by the
Madrid-Strasbourg group has been used.

\section{DSM application to Dark matter: Elastic scattering of LSP from 
$^{73}$Ge}

There is overwhelming evidence for the existence of dark matter in the universe
\cite{Jung-96,Debasish}. Up to now, the nature of this matter remains a mystery.
In recent years, there have been considerable theoretical and  experimental
efforts to detect the cold dark matter (CDM) which is thought to be the dominant
component of the dark matter \cite{Verg-2015}.  In the highly favored
Super Symmetric (SUSY) model, the most natural non-baryonic CDM candidate is the
lightest supersymmetric  particle (LSP) which is non-relativistic. 

Since the LSP (represented by $\chi$) interacts very weakly with matter,  its
detection is quite difficult. One possibility to detect LSP is through its
elastic scattering from nuclei. Inelastic channels are not excited since the
energy is too low to excite the nucleus and hence the cross section should be
negligible. On the other hand exotic WIMPs (weakly interacting massive
particles) can lead to large nucleon spin induced cross  sections which in turn
can lead to non-negligible probability for inelastic WIMP-nucleus scattering
\cite{Verg-2015}. Here we will consider only the  elastic channels. First we will
discuss briefly the formulation for LSP-nucleus scattering cross section
calculation and the related aspects of DSM. Next, results of the application to
$^{73}$Ge detector are described.

\begin{figure*}

\includegraphics[height=4in]{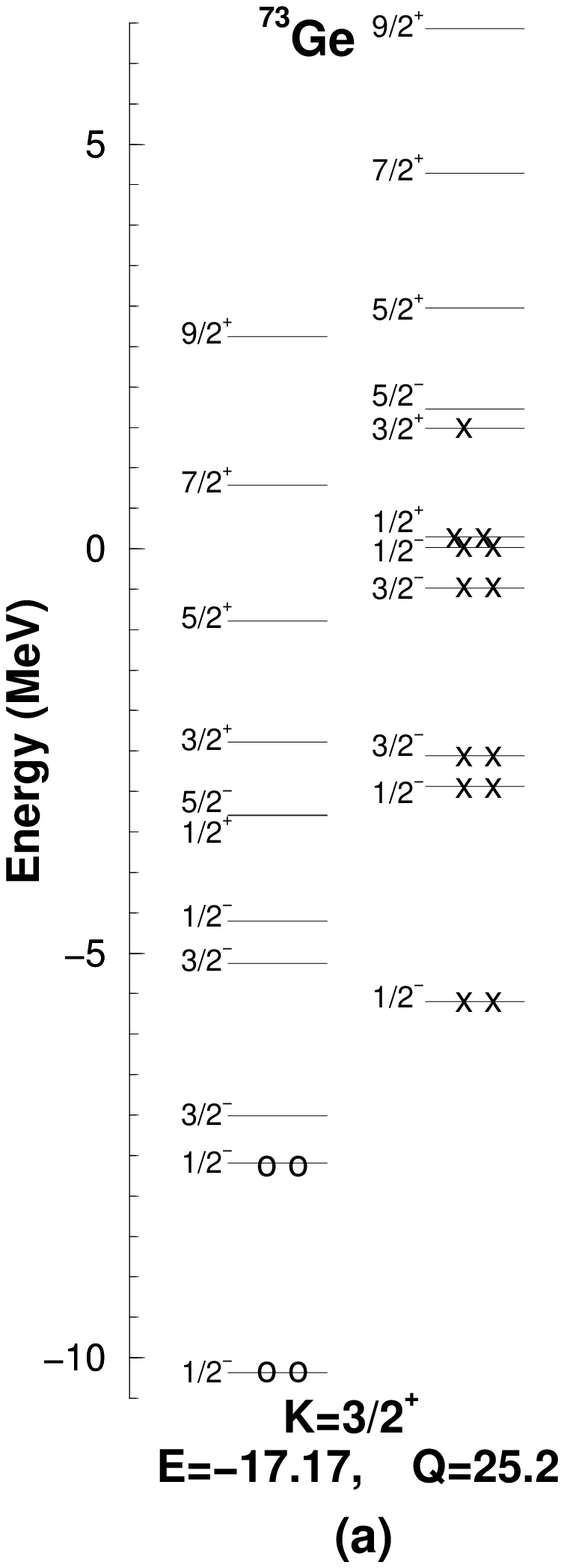} \vspace{8cm}
\includegraphics[height=4in]{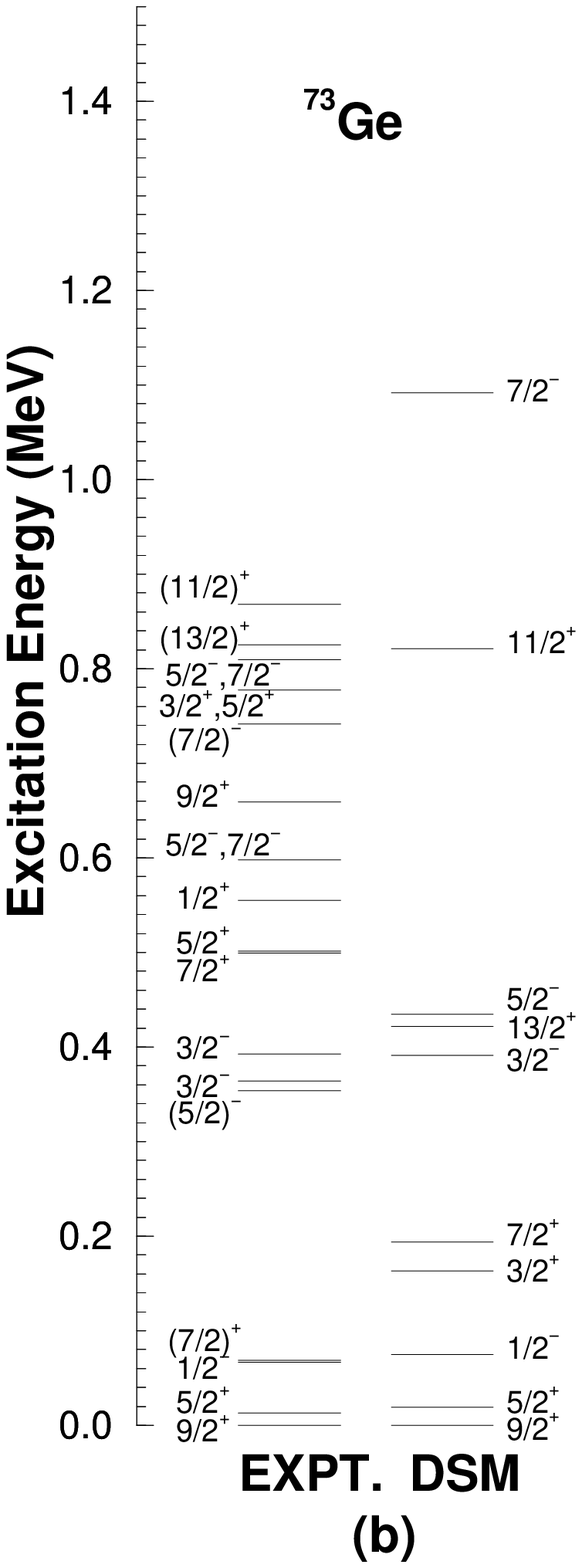}

\vskip -8cm
\caption{(a) The lowest prolate HF sp spectrum for $^{73}$Ge. The HF energy $E$
in MeV and the mass quadrupole moment Q in units of the square of the
oscillator length parameter $b$ are also given.   Protons are represented by
circles and neutrons by crosses. (b) Theoretical (DSM) and experimental (EXPT)
spectra of $^{73}$Ge. Data are taken from \cite{nndc}.}

\label{fig2}
\end{figure*}

\subsection{Formulation}

Defining the dimensionless quantity $u=q^2b^2/2=M_Ab^2\overline{Q}$ where $q$
represents the momentum transfer to the nuclear target, $b$ is the nuclear 
harmonic oscillator size parameter, $\overline{Q}$ is the energy transfer to the
nucleus and  $M_A$ is the nuclear mass, the LSP-nucleus differential cross
section in the laboratory frame is given by \cite{Holmlund-04, Divari-00},
\begin{equation}
\frac{d\sigma (u,v)}{du} = \frac{1}{2}\, \sigma_0\,\left(\frac{1}{m_pb}
\right)^2 \,\frac{c^2}{v^2} \,\frac{d\sigma_{AS}(u,v)}{du} \;;
\label{eq.4p5}
\end{equation}
\be
\barr{rcl}
\dis\frac{d\sigma_{AS}(u,v)}{du} & = & \l[f_A^0\Omega_0(0)\r]^2 F_{00}(u) +
2f_A^0 f_A^1 \Omega_0(0) \Omega_1(0) F_{01}(u)  \\
& &  + \l[f_A^1\Omega_1(0)\r]^2 F_{11}(u) + M^2 \;.
\earr \label{eq.4p6}
\ee
In the above, $m_p$ is the mass of the proton, $v$ is the LSP velocity with
respect to the earth and $\sigma_0=0.77 \times 10^{-38}$ cm$^2$. If the proton 
and neutron form factors $F_Z(u)$ and $F_N(u)$ are different, then
\be
M^2 = \l(f_S^0\, \l[Z F_Z(u) + N F_N(u) \r] + f_S^1\, \l[Z F_Z(u) - N F_N(u) 
\r]\r)^2 \;.
\label{eq.4p7}
\ee
Here, $f_A^0$ and $f_A^1$ represent isoscalar and isovector parts of the axial
vector current and similarly  $f_S^0$ and $f_S^1$ represent isoscalar and
isovector parts of the scalar current. These nucleonic current parameters depend
on the specific SUSY model employed. The spin structure functions
$F_{\rho\rho'}(u)$ with $\rho$, $\rho'$ = 0,1 are defined as
\be
\barr{l}
F_{\rho\rho'}(u) = \dis\sum_{\lambda,\kappa}\frac{\Omega_\rho^{(\lambda,\kappa)}(u)
\Omega_{\rho'}^{(\lambda,\kappa)}(u)}{\Omega_\rho(0)\Omega_{\rho'}(0)}\;;\\
\\
\Omega_\rho^{(\lambda,\kappa)}(u) = \sqrt{\frac{4\pi}{2J_i + 1}} \\
                \times \langle J_f \| \dis\sum_{j=1}^A \left[Y_\lambda(\Omega_j)
\otimes \sigma(j)\right]_\kappa j_\lambda(\sqrt{u}\,r_j) \times 
\omega_\rho(j) \|J_i\rangle
\earr \label{eq.4p8}
\ee 
with $\omega_0(j)=1$ and $\omega_1(j)=\tau(j)$; note that $\tau=+1$ for protons
and $-1$ for neutrons. Here $\Omega_j$ is the solid angle for the position
vector of the $j$-th nucleon and $j_\lambda$ is the spherical Bessel function. 
The static spin matrix elements are defined as $\Omega_\rho(0) =
\Omega_\rho^{(0,1)}(0)$. As has been described in  \cite{Holmlund-04},  the LSP
detection rate is given by the simple expression,
\be
R_0=8.9 \times 10^7 \times \frac{\sigma_{AS}(v_{esc})}{A\; m_{\chi}[GeV]
(m_pb)^2}[yr^{-1}kg^{-1}]\;.
\label{eq.4p9}
\ee
Note that $v_{esc} = 625$ km/s is the escape velocity of the LSP from the
milkyway and the LSP mass $m_\chi$ is taken to be $110$ GeV.  The
$\sigma_{AS}(v_{esc})$ is obtained using Eq. (2) and the Maxwell velocity
distribution (for $v$). In the integral over $u$, the lower limit involves the
detector threshold energy $Q$ and the upper limit involves $v_{esc}$.

The nuclear structure part is in the spin structure functions  and the form
factors. It is here DSM is used. The reduced matrix element appearing in Eq.
(\ref{eq.4p8}) can be evaluated in DSM. Here we need the sp matrix elements of
the operator of the form $t^{(l,s)J}_\nu$ and these are given by,
\be
\barr{l}
\langle n_il_ij_i\| \hat{t}^{(l,s)J}\| n_kl_kj_k \rangle = 
\dis\sqrt{(2j_k+1)(2j_i+1)(2J+1)(s+1)(s+2)} \\
\\
\left\{\begin{array}{ccc}
l_i & 1/2 & j_i\\
l_k & 1/2 & j_k\\
l   & s   & J
\end{array}\right\} \;
\langle l_i \| \sqrt{4\pi}Y^l \|l_k \rangle \;\langle n_il_i \|j_l(kr)\| 
n_kl_k \rangle \;.
\earr \label{eq.4p12}
\ee
In the above equation, $\{--\}$ is the nine-$j$ symbol. 

\subsection{Results and discussion}

Above formulation is used for LSP detection rates for scattering from $^{73}$Ge
with DSM for the nuclear structure part. The sp orbits employed are 
$^2p_{3/2}$, $^1f_{5/2}$,  $^2p_{1/2}$ and $^1g_{9/2}$ with$^{56}$Ni core and
the sp energies are taken as  0.0, 0.78, 1.08 and 4.90 MeV respectively. The
effective interaction used is the modified Kuo interaction \cite{Kuo-M}.  The HF
sp spectrum is shown in Fig. \ref{fig2}a. For $^{73}$Ge, the experimental energy
spectrum has positive and negative parity levels at low energy. Hence, for band
mixing in DSM  three intrinsic states with positive parity and three with
negative parity are considered. The final energy spectrum and its comparison
with experiment is shown  in Fig.  \ref{fig2}b. Since  spin contributions play
an important role in the the calculation of  the decay rates, the magnetic
moment is decomposed into orbital and spin parts for this nucleus.  The DSM
value for the magnetic moment (using bare values for the $g$-factors) is 
$-0.811 \;\mu_N$ and it is close to experimental value $-0.879\;\mu_N$
\cite{nndc}.  The matrix elements of the proton orbital and spin angular momenta
are $0.581$ and $-0.001$ respectively and similarly, for neutron  the values are
$3.558$ and $0.362$ respectively.

\begin{figure*}
\includegraphics[height=3in]{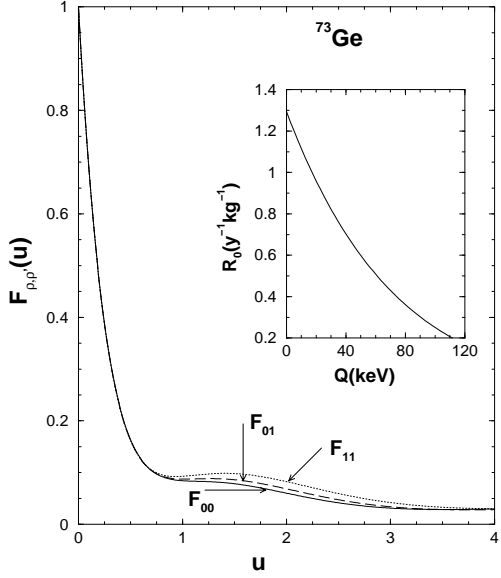} 
\caption{Spin structure functions for $^{73}$Ge as a function of momentum 
transfer $u$. Shown in the insect figure is LSP detection rate as a function of
$Q$, the detector threshold energy.}
\label{fig3}
\end{figure*}

Depending on the SUSY parameters, the detection rate varies widely as described
in \cite{Holmlund-04}. The same feature is also found in DSM. The values of the
parameters $f_A^0$, $f_A^1$, $f_S^0$ and $f_S^1$ are taken from \cite{Verg-1996}
and they are $3.55 \times 10^{-2}$, $5.31 \times 10^{-2}$, $8.02 \times 10^{-4}$
and $-0.15 \times f_S^0$ respectively. For $^{73}$Ge, DSM gives the values of
$\Omega_0$ and $\Omega_1$ to be  $0.798$ and $-0.803$. These values are smaller
than those quoted in \cite{Holmlund-04}, where a quasi-particle-phonon model
(QPM) is used, by 20 to 30 percent.  The spin structure functions which do not
depend on the oscillator length parameter are plotted in Fig. \ref{fig3}. The
structure functions for $^{73}$Ge are similar to those obtained using QPM in 
\cite{Holmlund-04}. Ressell {\it et al.} \cite{Resell-1993} calculated
$S_{\rho\rho^\prime}$, that  are related to the spin structure functions defined
above, using shell model.  The spin structure functions from DSM are similar to
their values.  Following these, the detection rate as a function of $Q$ is
obtained using Eq. (\ref{eq.4p9}) and the results are shown in Fig
\ref{fig3}. Results in the figure show that $^{73}$Ge is a good detector for 
detecting dark matter.

\section{Conclusions}

Applications of DSM to two current problems of interest in nuclear structure are
presented in this paper. They are: (i) analysis of recent data on low-lying
$T=1$ and $T=0$ bands in the heavy N=Z nucleus $^{66}$As; (ii) detections rates
for the light supersymmetric particle, a dark matter candidate, with $^{73}$Ge
as the detector. In future heavy N=Z nuclei will be further analyzed to bring 
out isoscalar pairing vs isovector pairing in these nuclei by using extended
pairing plus quadrupole-quadrupole Hamiltonian of the Sofia group \cite{Kalin}
and comparing the results of DSM obtained with this interaction with those using
realistic interactions. In addition, in the topic of dark matter, DSM will be
employed to study inelastic (spin dependent) WIMP-nucleus scattering in
$^{83}$Kr and this is unlike LSP that involves only elastic scattering.

\section*{Acknowledgments}

The authors are thankful to P.C. Srivastava for useful correspondence.  R. Sahu
is thankful to SERB of Department of Science and Technology (Government of
India) for financial support.

\end{document}